\documentclass{elsart}
\usepackage{graphicx}

\begin{document}
\begin{frontmatter}
\title {Precision Measurement of 
Energy and Position Resolutions of the BTeV Electromagnetic
Calorimeter Prototype}
\author[IHEP]{V.A.~Batarin},
\author[FNAL]{T.~Brennan},
\author[FNAL]{J.~Butler},
\author[FNAL]{H.~Cheung},
\author[IHEP]{A.A.~Derevschikov},
\author[IHEP]{Y.V.~Fomin},
\author[UMN]{V.~Frolov},
\author[IHEP]{Y.M.~Goncharenko},
\author[IHEP]{V.N.~Grishin},
\author[IHEP]{V.A.~Kachanov},
\author[IHEP]{V.Y.~Khodyrev},
\author[SYR]{K.~Khroustalev},
\author[IHEP]{A.S.~Konstantinov},
\author[IHEP]{V.I.~Kravtsov},
\author[UMN]{Y.~Kubota},
\author[IHEP]{V.M.~Leontiev},
\author[IHEP]{V.A.~Maisheev},
\author[IHEP]{Y.A.~Matulenko},
\author[IHEP]{Y.M.~Melnick},
\author[IHEP]{A.P.~Meschanin},
\author[IHEP]{N.E.~Mikhalin},
\author[IHEP]{N.G.~Minaev},
\author[IHEP]{V.V.~Mochalov},
\author[IHEP]{D.A.~Morozov},
\author[SYR]{R.~Mountain},
\author[IHEP]{L.V.~Nogach},
\author[IHEP]{A.V.~Ryazantsev},
\author[IHEP]{P.A.~Semenov\thanksref{addr}},
\thanks[addr]{corresponding author, email: semenov@mx.ihep.su}
\author[IHEP]{K.E.~Shestermanov},
\author[IHEP]{L.F.~Soloviev},
\author[IHEP]{V.L.~Solovianov\thanksref{deceased}},
\thanks[deceased]{deceased}
\author[SYR]{S.~Stone},
\author[IHEP]{M.N.~Ukhanov},
\author[IHEP]{A.V.~Uzunian},
\author[IHEP]{A.N.~Vasiliev},
\author[IHEP]{A.E.~Yakutin},
\author[FNAL]{J.~Yarba},
\collab{BTeV electromagnetic calorimeter group}
\date{\today}

\address[IHEP]{Institute for High Energy Physics, Protvino, Russia}
\address[FNAL]{Fermilab, Batavia, IL 60510, U.S.A.}
\address[UMN]{University of Minnesota, Minneapolis, MN 55455, U.S.A.}
\address[SYR]{Syracuse University, Syracuse, NY 13244-1130, U.S.A.}

\begin{abstract}
     The energy dependence of the energy and position resolutions
of the electromagnetic calorimeter prototype made of lead tungstate 
crystals produced in Bogoroditsk (Russia) and
Shanghai (China) is presented.  These measurements were carried out
at the Protvino accelerator using a 1 to 45 GeV electron beam. The crystals were coupled to photomultiplier tubes. The dependence of energy and
position resolutions on different factors as well as the measured electromagnetic shower lateral profile are presented.
\end{abstract}
\end{frontmatter}

\section{Introduction}
     BTeV is a dedicated experiment at the Tevatron proton-antiproton
collider at Fermilab that will study $b$ and $c$ quark decays \cite{prop}.
A thorough investigation of B decays requires the ability to 
study decay modes containing single photons, $\pi^0$'s, and $\eta$'s. 
Total absorption shower counters made of scintillating crystals 
have been known for decades for their superb energy and spatial 
resolutions. The crystals act as both the shower
development medium and light producer.
Lead tungstate (PbWO$_4$) crystals are distinguished 
with their high 
density, short radiation length and small Moliere radius. The 
CMS and ALICE experiments at the CERN LHC have chosen these crystals 
for their 
Electromagnetic calorimeters \cite{cms_tdr},\cite{alice_tdr}. 
 An electromagnetic 
calorimeter made of PbWO$_4$ crystals has also been selected as 
the baseline for the BTeV experiment. Unlike CMS or ALICE BTeV ECAL is 
not in a high magnetic field, so we can use photomultiplier 
tubes rather than avalanche photodiodes or vacuum phototriodes. 
This should provide less noise and better resolution at low energies. 
According to the PbWO$_4$ manufacturers specifications the expected 
light collection is $\approx$5000 photoelectrons at 1 GeV. 

     The performance of lead tungstate crystals produced in 
Bogoroditsk Techno-Chemical Plant (BTCP) and Shanghai Institute
of Ceramics (SIC) has been 
studied at the accelerator U70 in Protvino, Russia in 2001.
The specific goals were to understand how to set specifications    
for crystal production and measure the predicted energy and position
resolution.

\section{Testbeam Facility}

\begin{figure}[b]
\centering
\includegraphics[width=0.95\textwidth]{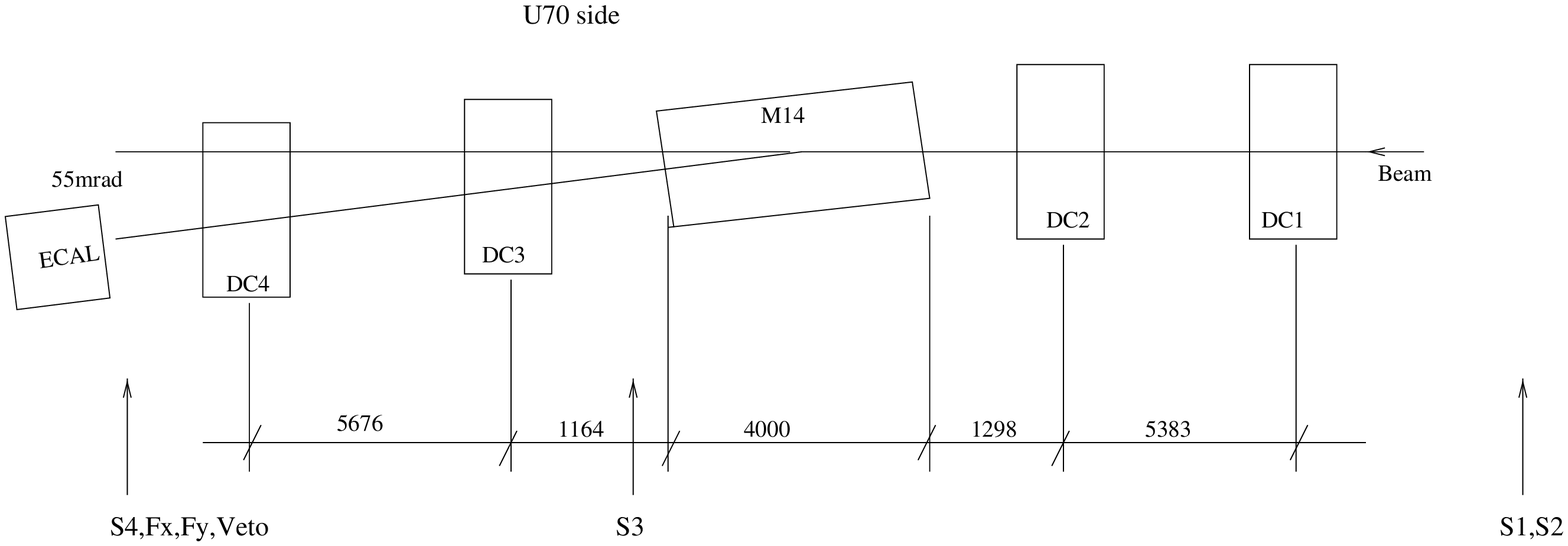}
\caption{Testbeam setup. DC indicates sets of drift chambers,
while M14 is a dipole bending magnet and ECAL is a calorimeter 
prototype. (All distances are in mm.)}
\label{fig:beam_setup}
\end{figure}

    The BTeV ECAL testbeam setup consisted of a $5\times5$ array of PbWO$_4$ 
crystals coupled to ten-stage 
photomultiplier tubes, a beam with a momentum tagging system on individual 
particles and a trigger system using scintillation counters (see 
Fig.~\ref{fig:beam_setup}). The crystals had a square cross-section 
of 27$\times$27~mm$^2$ and were 220 mm long and  wrapped with 
170$\mu$m Tyvec.

PbWO$_4$ light yield strongly depends on 
temperature \cite{cms_tdr},\cite{pwo_first}. 
To eliminate the effects of temperature variation 
crystals were placed inside a thermally insulated
light-tight box. We controlled the temperature in the box using a 
LAUDA cryothermostat with an accuracy of $\pm 0.1$~$^\circ$C. To measure the
temperature of the crystals, 24 separate temperature 
sensors were placed on the front and rear ends of various crystals. 

A moving platform was 
used to inject the beam on different points of the crystal matrix. 
The platform was 
designed to move in the both vertical and horizontal directions and 
was controlled remotely. The box was placed 
on a revolving table mounted on the platform. Rotation of the table 
was important for studying an angular resolution and optical 
transmission of the crystals.

    In order to measure the energy resolution we need to take into 
account the electron beam energy spread which was 2-3\% at high 
energies up to 45 GeV and was 5-7\% at low energies down to 1 GeV.
These values are significantly larger than expected performance of the
crystals. Thus we constructed a system that measured the momentum of each 
electron in the beam \cite{nim1}. It consisted
of four drift chamber stations and an analyzing magnet. 
Three stations had X and Y chamber pairs, and the fourth one
had only an X chamber pair. Each chamber covered an area 
about $20\times20$ cm$^2$. 
A gas mixture
of 70\% Ar and 30\% isobutane was used. 
The position resolution of the chambers was 160~$\mu$m.
An electron momentum was measured with a resolution of 0.13\% at 45 GeV
where the multiple scattering was negligible. At 1 GeV the
resolution of 2\% was caused mainly by 
multiple scattering of electrons on materials along the 
beam line (the flanges of the vacuum tubes and drift chambers,
0.006 of radiation length in total.)

    The positions of the beam telescope scintillating counters 
S1, S2, S3 and S4 are shown in Fig.~\ref{fig:beam_setup}. 
Counters sizes were 10~cm in diameter for S1, S2 and S3 and $15\times 15$~cm
for S4.
The coincidence of the signals from these counters formed 
the main beam trigger. The finger counters F$_x$
and F$_y$ with sizes 5~mm were used mainly in calibration and 
precise energy resolution measurements to limit the beam size. 

\begin{figure}
\centering
\includegraphics[width=0.95\textwidth]{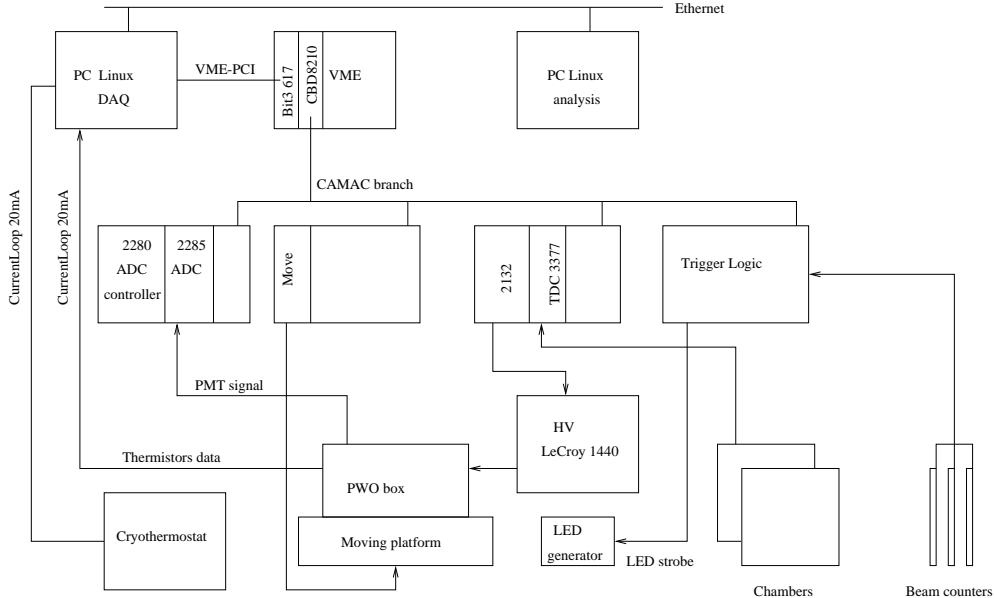}
\caption{Testbeam DAQ diagram}
\label{fig:daq}
\end{figure}

Data acquisition (DAQ) was CAMAC-based and PC-controlled using VME
(see Fig.~\ref{fig:daq}).  The DAQ system consisted of 
\begin{itemize}
\item CAMAC crate with an ADC system for crystals readout,
\item TDC modules measuring the timing of  drift chambers signals, and 
\item trigger  logic.
\end{itemize}
A CES CBD8210 CAMAC branch driver VME board was used as an interface 
to the CAMAC crates. It also provided interrupts to control DAQ 
program flow. The DAQ programs ran in a PC under Linux operating system. 
A Bit3 PCI-VME interface was used to communicate
with the VME crate. The Fermilab software tool HistoScope was used both as 
run-control frame
and for a simple on-line analysis of the data. Data also were sent
to another computer for more detailed analysis and archiving. This was 
implemented by pipe and TCP/IP socket system calls.
The slow control subsystem included 
\begin{itemize}
\item high voltage control and monitoring by LeCroy 1440 HV power supply,
\item temperature control of the crystal's box using the LAUDA cryothermostat,
\item monitoring of temperatures in the crystal matrix by a set of thermistors,
\item monitoring of the phototubes gains via pulsed LED, 
\item the moving platform control elements.     
\end{itemize}

  Light from each crystal was viewed by a PMT through an optical 
grease coupling. We used
R5800 Hamamatsu phototubes equipped with transistorized bases 
developed at Fermilab.
High voltage to the tubes was supplied by a LeCroy 1440 HV system. 
Signals were sent to the control 
room patch-panel without any connection to ground inside the crystal 
box to avoid ground loops. LeCroy 2285 15-bits integrating ADC modules 
were  used to measure the signal
charge within 150 ns gate, chosen to be somewhat 
longer than the natural decay
time of the crystals, where we expect 99\% of the light in 100 ns. 
The ADC sensitivity could be programmed with a power controller module.
During most of our studies we used 30 fC per count. 

   Signals from drift chamber sense wires were sent to 
amplifier-discriminator cards that were installed on the chambers. 
We used UPD16 cards produced by the IHEP Electronics 
department with a $2.5~{\mu}A$ threshold. The output from the UPD16  
was a balanced ECL signal with an 80 ns width. The output signal went
to a TDC input using a 60-meter twisted-pair cable and started the TDC
conversion operation. The conversion was stopped by a delayed
trigger. The delay was chosen to have all start--stop time 
values in the 1 ms timeout window. With a 10-bit
TDC scale, the precision is 1 ns per count.

\section{Monitoring System}

One of the problems in high precision measurements
is the long-term stability of measuring devices. Placing the
crystal matrix together with photomultipliers into a 
thermoinsulated box we have reduced already the influence 
of the temperature, which is  the most important instability factor.
But we still had some electronics outside the box. Besides, some other 
factors (such as photomultiplier gain dependence on mean anode current)
might cause drift in the output signal. To monitor and take
into account possible drift, a LED pulser system was implemented. 
A generator fired the super-bright blue LED 
with 470 nm wavelength, that provided light to 50 fibers. Each 
fiber was fed to one (crystal+PMT+ADC) channel 
of the setup. To monitor the stability of the LED pulser 
we used 2 PIN-diodes HAMAMTSU S6468-05 as well as a dedicated PMT 
with calibrated 
light from a light source with YAP:Ce crystal \cite{yap}.  
The LED pulses had the height 2--3 times less and 
the width similar  to those of the scintillation light 
produced in the PbWO$_4$ crystal by 27 GeV electron. The
LED was triggered by the DAQ system 10 times during each of the 
intervals between accelerator spills. This provided enough 
statistics to accurately monitor the stability 
every few minutes. We had checked the stability of the 
LED pulser relative to the YAP:Ce calibrated light  and found it 
was better 0.1\% during at least 15 hours.

\section{Crystal Light Response Uniformity}

    GEANT simulations show that an adequate light response 
uniformity along the length of the crystal is a key to achieve 
excellent energy resolution (see Fig.~\ref{fig:mc_eres}). 
The non-uniformity of the light yield (LY) along the crystal length
contributes to the constant term of the relative energy
resolution. 

    To measure the LY uniformity, the $5\times5$ crystal matrix was rotated
by 90$^{\circ}$ around the vertical axis and crystals were
scanned using a muon beam in 1 cm steps. 
The position of the muon track going through the crystal was 
reconstructed using the drift chambers. Pulse-height distribution 
collected for
each of the 1 cm intervals along the crystal lengths were fitted with 
a modified Landau distribution to obtain a peak position. 

\begin{figure}
\centering
\includegraphics[width=0.95\textwidth]{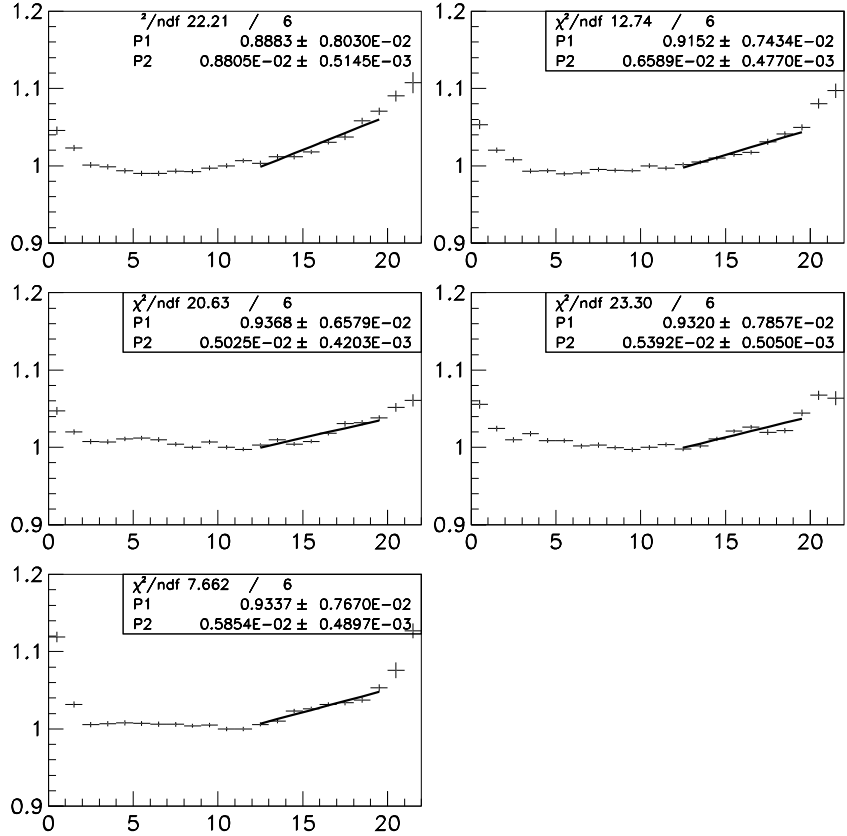}
\caption{Fit results for the energy loss distributions of 5 crystals
as a function of the position along the crystal. PMT position is at the
X=0~cm. LY on $Y$-axis were normalized to the LY at X=11~cm. Each plot 
corresponds to one crystal. 
}
\label{fig:unif_fit}
\end{figure}

    The peak position of the energy loss distribution for minimum 
ionizing particle as a function of the distance to the PMT is 
shown in Fig.~\ref{fig:unif_fit}. The PMT position is at $X=0$~cm. The data 
were fitted in the region of the expected shower maximum 
(3 to 10 X$_0$) to a straight line in order to determine the 
slope of the LY uniformity. The LY values were normalized to the 
value of LY at $X=11$~cm. 

\begin{figure}
\centering
\includegraphics[width=0.7\textwidth]{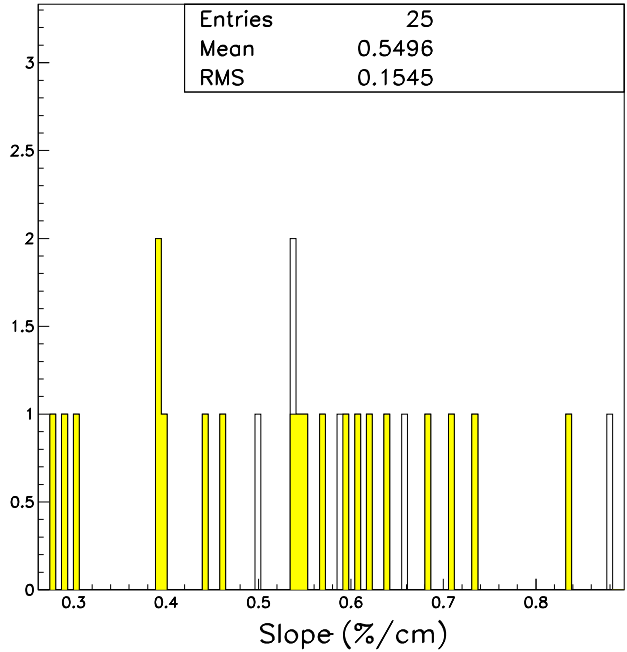}
\caption{The distribution of the LY uniformity slope.}
\label{fig:unif_hist}
\end{figure}

     A distribution of the slopes of the  LY uniformity was obtained for the groups 
of 25 crystals each from Bogoroditsk and Shanghai. The results are shown in 
Fig.~\ref{fig:unif_hist}. 
No difference between the Bogoroditsk and the Shanghai crystals 
was observed.

\section{Energy Resolution and Comparison With GEANT Simulations}
       
Using the test beam setup described above the energy resolution of the
electromagnetic calorimeter prototype ($5 \times 5$ matrix) were measured. 
The electron beam was directed to the center of the matrix. 
To disentangle various contributions to
the energy and position resolution six electron beam energies
of 1, 2, 5, 10, 27, and 45 GeV were used. 

\begin{figure}[p]
\centering
\includegraphics[width=0.95\textwidth]{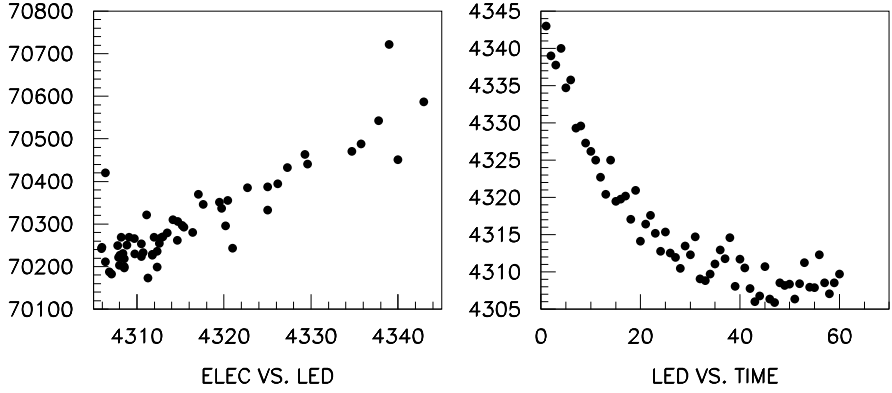}
\caption{(left) The correlation between the LED signal (on the $X$ axis) and 
$5\times 5$ energy sum for electrons (on the $Y$-axis). (right) 
Dependence of the LED signal in ADC counts vs the time
(on the $X$ axis). Each point is an average
over 90 seconds. Note, the highly suppressed zero on the scales.
(All data are at 27 GeV.) 
}
\label{fig:gain_var}
\end{figure}

\begin{figure}[p]
\centering
\includegraphics[width=0.8\textwidth]{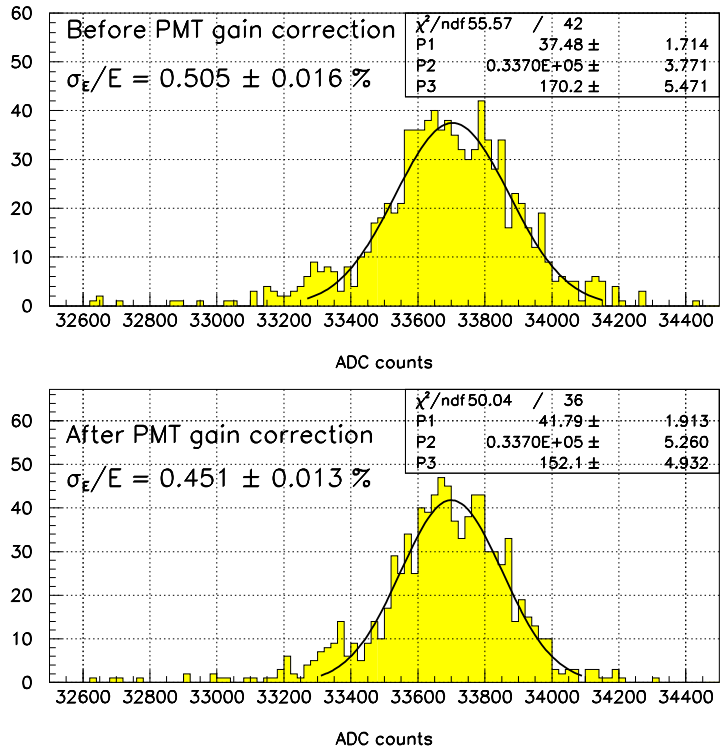}
\caption{Energy resolution before and after the PMT gain correction 
using the LED pulser information. Data are collected 
using a 27-GeV electron beam.
}
\label{fig:led_cor}
\end{figure}

To measure a good energy resolution of the calorimeter prototype
we had to take into account a photomultiplier gain 
instability. Variations of the PMT gain were caused by 
the changes of the electron beam intensity. To monitor PMT 
gain we used LED pulser monitor system described above. Monitor
system data were used to correct PMT gain.
Fig.~\ref{fig:gain_var} (right),  shows the average pulse 
heights from LED pulser signals as a function of time. 
The left plot
in Fig.~\ref{fig:gain_var} shows a linear correlation between the 
average LED signal and electron beam pulse height. 
The energy resolution before and after the PMT gain 
correction with the use of the LED data is presented in 
Fig.~\ref{fig:led_cor}. All results below were obtained
with PMT gain correction.

\begin{figure}[b]
\centering
\includegraphics[width=0.95\textwidth]{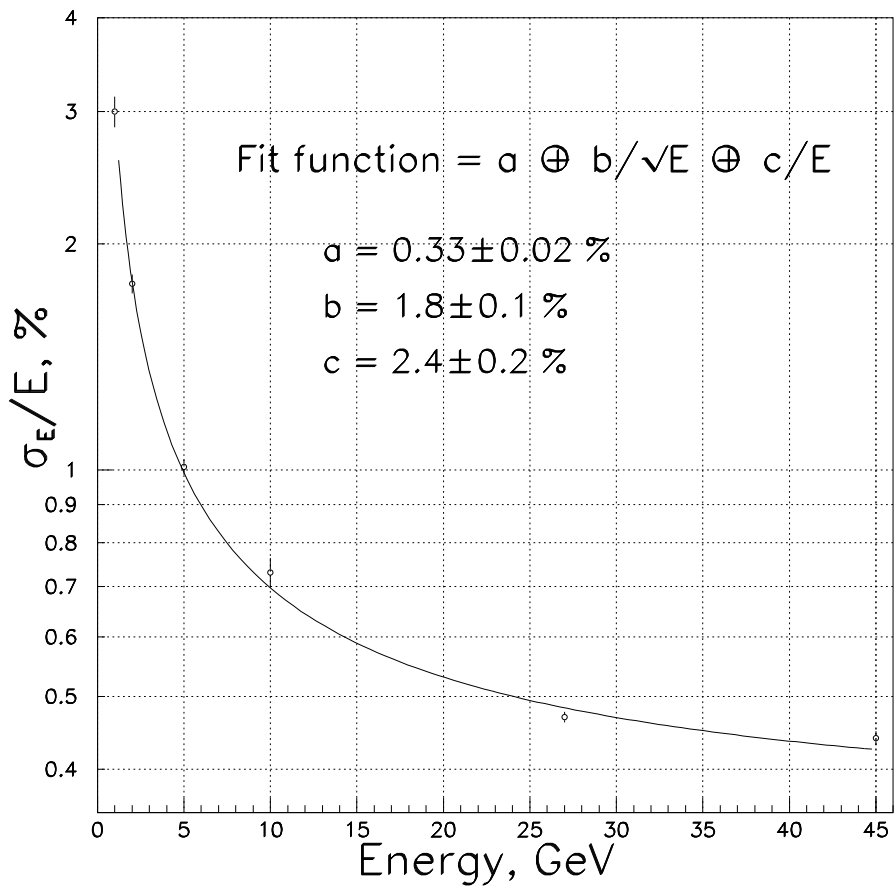}
\caption{Measured energy resolution of the $5\times 5$ crystal matrix. 
A curve shows experimental data fit result.}
\label{fig:e_res}
\end{figure}

      The energy resolution $\sigma_E/E$ as a function of $E$ is shown
in Fig.~\ref{fig:e_res}.
The energy resolution is described well using a function
\begin{equation}
\sigma_E/E = a \oplus b/\sqrt{E}  \oplus c/E~~[\%],
\end{equation}
where $E$ is in GeV, {\it a} represents a constant term 
arising from calibration errors, shower leakage,
mostly from the back of the crystals, and non-uniformity in the
light collection efficiency along the length of the crystals.
The stochastic term, $b = (1.8 \pm 0.1)$\%, arises from photon
statistics and leakage of shower, mainly in the transverse directions
outside the 5$\times$5 crystal array. The last term $c = (2.4 \pm 0.2)$\%,
sometimes called a noise term, usually arises from noise of the 
photon detection
electronics, which was negligible in our case. Instead, the momentum 
measurement resolution arising from multiple scattering of the 
electrons in the 
beam line is estimated to contribute 2.2\% to this term and is, in fact,
consistent with what we observe.

      The measured constant term is  $a = (0.33 \pm 0.02)$\%.
Our Monte Carlo studies show that the shower fluctuation and 
non-uniformity contribute 0.23\% and 0.27\%, respectively. The measured
longitudinal non-uniformities were used as inputs in this Monte Carlo
study. Adding these contributions in quadrature we estimate the constant 
term to be 0.35\%, which is consistent with what we measure.

      The same Monte Carlo studies show that shower fluctuation
results in 0.72\% contribution to {\it b}. To estimate the other major
contributions in the {\it b} term we need to know the photo-electron yield.
The vendors of the crystals BTCP (Bogoroditsk Techno-Chemical Plant),
Russia and SIC (Shanghai Institute of Ceramics), China measured
this number to be about 10 pe/MeV using Cs$^{137}$ and Co$^{60}$ gamma
sources and 2" PMT's with bialkali photo cathode, covering the
entire crystal end. Since the PMT's used in the beam test have
sensitive areas of ($22 \pm 1$) mm diameter attached to the crystal
ends measuring 27 mm square, the covered area is only ($52 \pm 5$)\%.
This implies that photo-electron yield in our beam test studies
is 5 pe/MeV, and its contribution in the {\it b} term is 
($1.45 \pm 0.07$)\%. Combining these two contributions as well as
an additional small contribution from the LY non-uniformity 
to the {\it b} term, we
expect the stochastic term to be ($1.68 \pm 0.07$)\%,
which is consistent with the measured value of ($1.8 \pm 0.1$)\%. 
We did not see big difference in
energy resolution for crystals produced in Bogoroditsk and Shanghai. 

\begin{figure}
\centering
\includegraphics[height=0.75\textwidth]{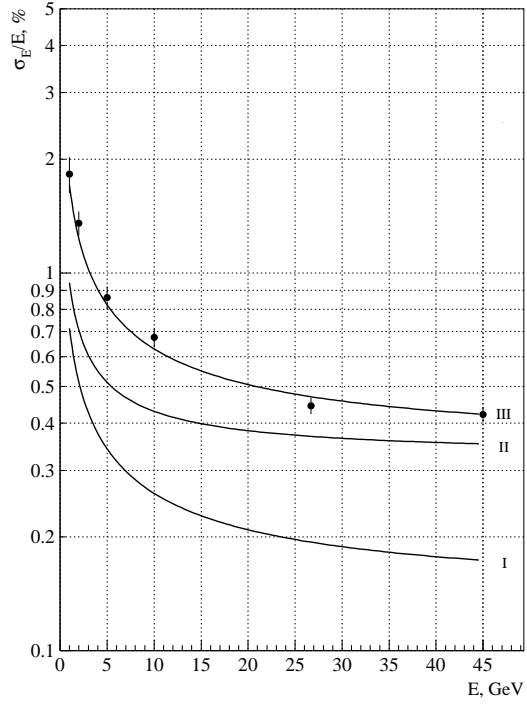}
\caption{Comparison of the measured energy resolution with GEANT simulations.
Curve I shows Monte Carlo result for shower fluctuations. 
In curve II light yield non-uniformity along the crystal 
is taken into account. In curve III  photon statistics is included.
Dots represent the experimental results. 
}
\label{fig:mc_eres}
\end{figure}

The results of GEANT version 3.21 \cite{geant} Monte Carlo predictions 
are compared with the data in Fig.~\ref{fig:mc_eres}. 
\begin{figure}
\centering
\includegraphics[width=0.55\textwidth]{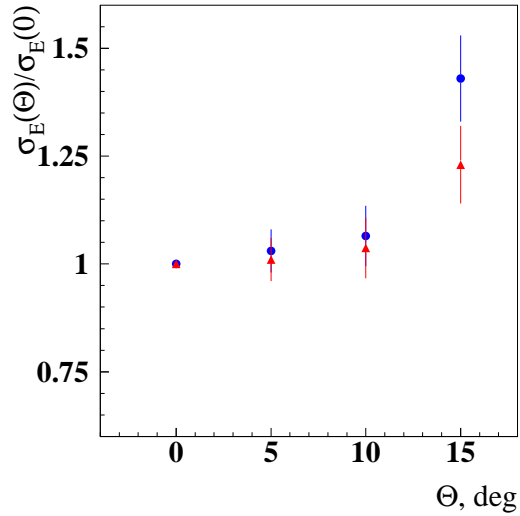}
\caption{Energy resolution dependence on the angle of electron incidence.
Resolution along the Y-axis normalized to the resolution at 0$^\circ$. 
Dots represent results from 27 GeV electron beam, 
triangles -- 10 GeV electrons 
}
\label{fig:e_angle}
\end{figure}
Curve I includes the effects of shower 
fluctuation and leakages for our particular crystal size. 
The light yield non-uniformity along 
the crystal contribution is added on curve II. 
Photon statistics is included into the calculations in 
curve III. Dots in the Fig.~\ref{fig:mc_eres} represent 
the experimental data which had have the effects of multiple 
scattering of the beam electrons and tagging system resolution 
removed. 

       We studied the dependence of the energy resolution on the
angle of electron incidence on the crystals.
No changes were observed up to a few degrees. 
In  the 5$\times$5 crystal array it began to deteriorate 
at angles greater than 5$^\circ$ (see Fig.~\ref{fig:e_angle}).
The effect is more-or-less constant over the energy range we studied. 

\section{Temperature Dependence of the Crystal Light Output}

\begin{figure}
\begin{tabular}{lr}
\includegraphics[width=0.5\textwidth]{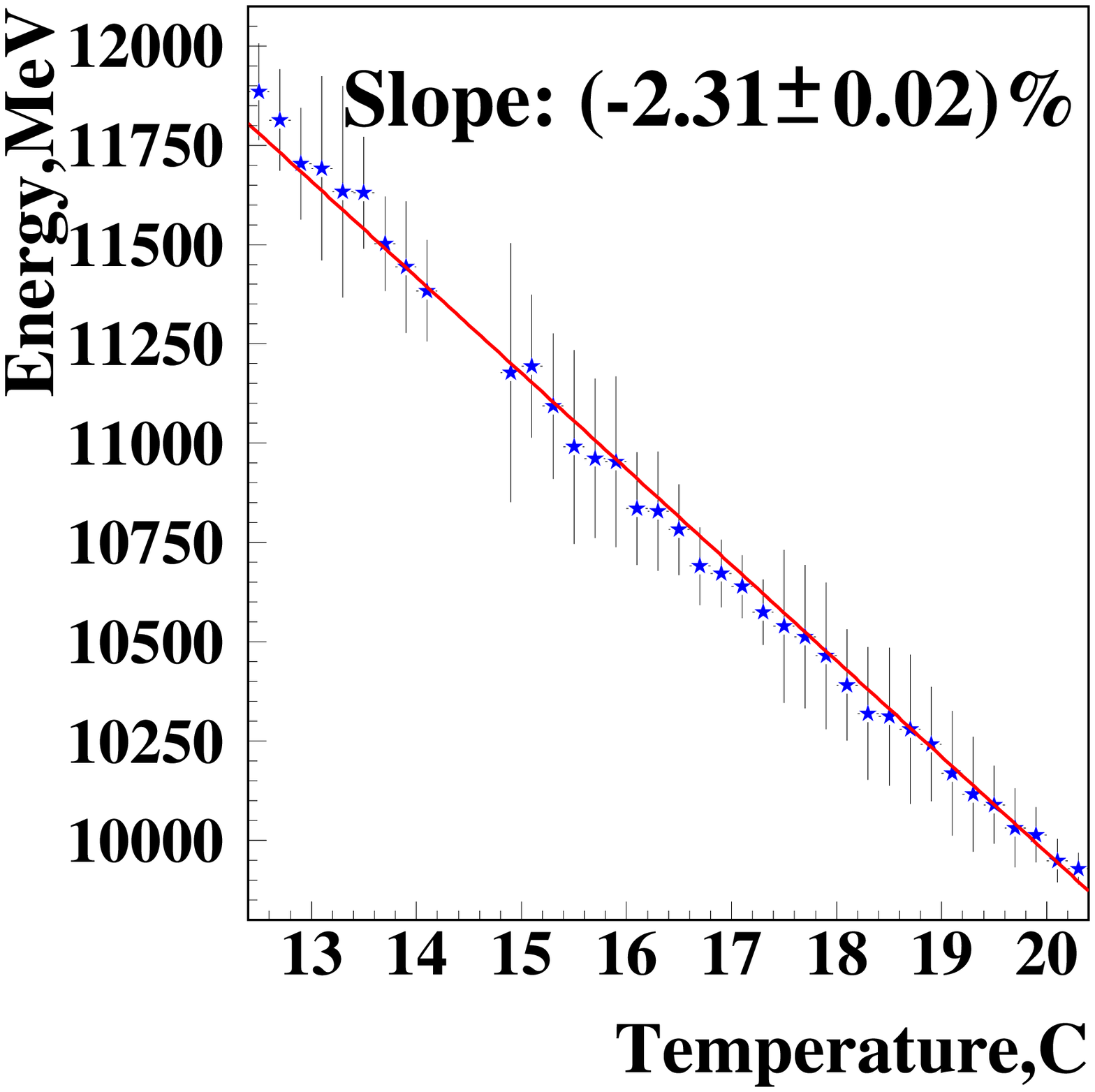} &
\includegraphics[width=0.5\textwidth]{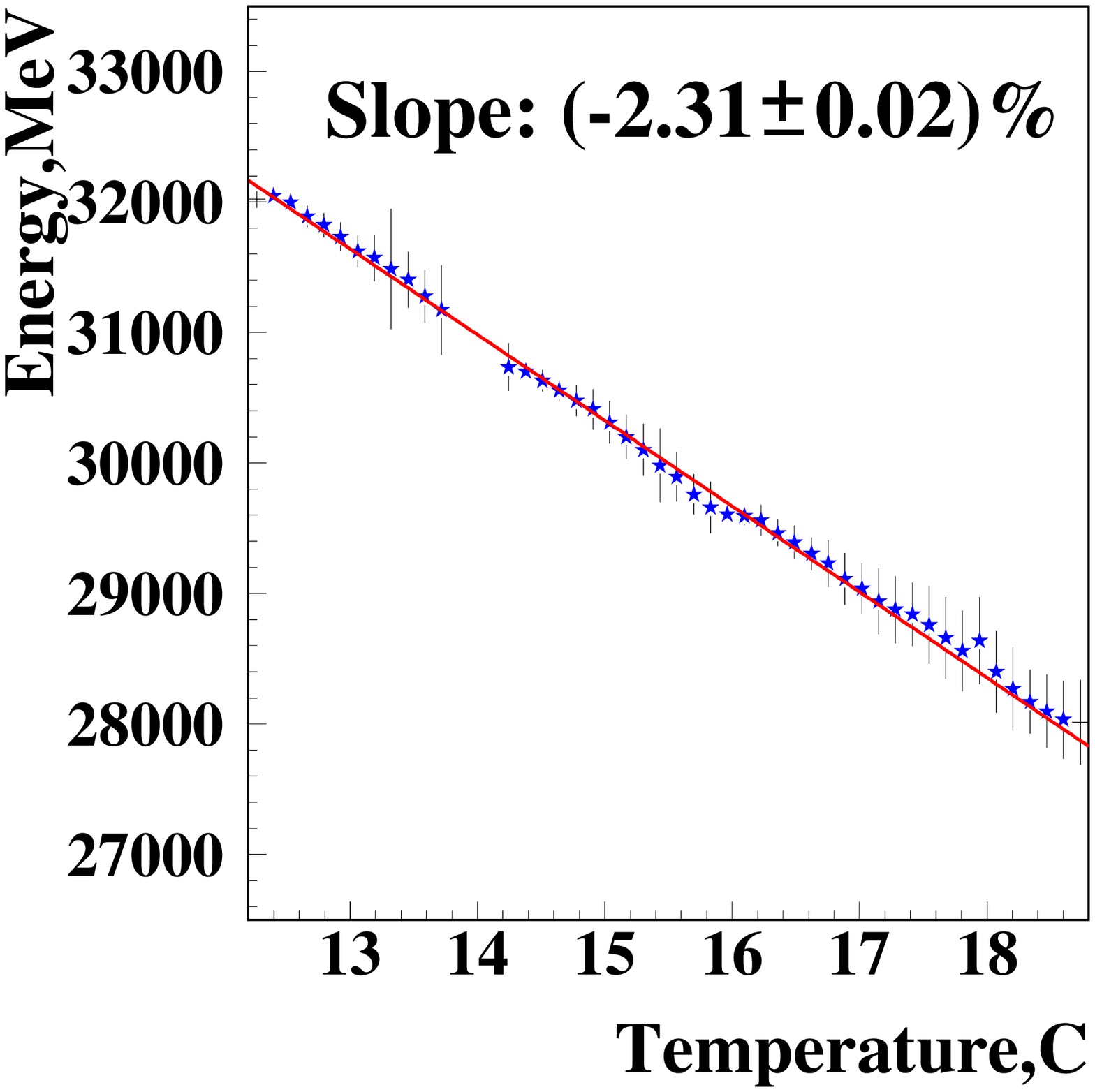}\\
\end{tabular}
\caption{Temperature dependence of light output for one of the crystals. 
Temperature in $^{\circ}$C is along the {\it X}-axis.  
Light output in terms of energy is along the {\it Y}-axis.
Measurements were done with 10 GeV (left) and 27 GeV(right) electrons. 
}
\label{fig:tempr}
\end{figure}

       We made an independent measurement of a temperature variation 
effect on the PbWO$_4$ crystal light output. The temperature dependence
measurements were made at electron energies of 10 and
27 GeV. The rate of the temperature change was about 1~$^{\circ}$C/hour
during both the warm up and the cool down periods. The temperature
inside the box was measured using our 24 thermistors array, averaging
the values. The temperature was measured once per spill
(approximately 0.1 Hz). The slope of the change in the vicinity of
 18~$^{\circ}$C was found to be about -2\% per $^{\circ}$C,
in agreement with previous measurements. Fig.~\ref{fig:tempr} shows
our measurements with 10 and 27 GeV electrons.

\section{Electromagnetic Shower Lateral Profile}

      We have measured the transverse electromagnetic shower profile 
and also its dependence on energy. 
We use only events 
with an electron hitting the central crystal of the array
for the further analysis to avoid shower leakage outside matrix.

      In this analysis, the crystals were divided into virtual
squares of 1 mm$^2$. During the accumulation of the data for 
shower profiles three two-dimensional arrays 135$\times$135 were used, where
the two dimensions stood for {\it x} and {\it y} coordinates perpendicular 
to the beam. Each cell of the arrays corresponded to the 1~mm$^2$ of the 
crystal matrix. Each of the arrays contained 
number of events, the energy sum, and the sum of the energy squared 
correspondingly.
For any selected event only 25 cells in each 
of these arrays were used for accumulation. Coordinates of each cell to be 
filled were determined by the distances {\it x} and {\it y} of 
the center of each 
out of 25 crystals in the matrix relative to the 
electron coordinates determined by the drift chambers. 

After
the accumulation of the data was done, the arrays were modified.
Considering a symmetry of the shower profile in the both 
projections, only 1/8 of the full arrays are left. Thus we mapped the 
information from the cells symmetrical relative to the center 
of the 5$\times$5 matrix. Finally three ``triangular''
tables (the 5$\times$5 crystal matrix viewed
from the center of the matrix appears as a triangle with 68 cells along 
the two equal sides, that is along 2.5 crystals 
with a step of 1 mm) were determined. The first one contained 
the number of events accumulated for each cell, the second one contained  
an average energy deposit in particular cell, and the third one contained 
the r.m.s. ($\sigma$) of this average energy deposit. Each final table had 
2346 elements. These tables were accumulated for the energies 
1, 2, 5, 10, 27, and 45 GeV. The energy sharing information is  important for developing 
an algorithm of a reconstruction of two gamma-quanta from two 
strongly overlapped showers at different energies.       

\begin{figure}
\centering
\includegraphics[width=0.95\textwidth]{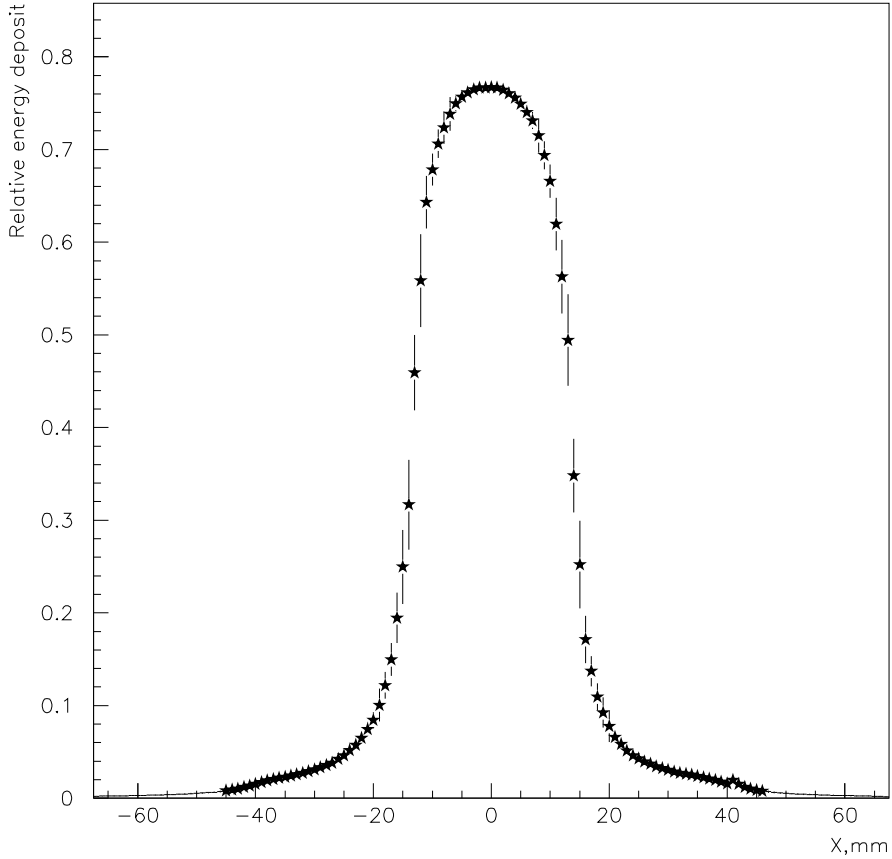}
\caption{Electromagnetic shower lateral profile for 45 GeV electrons.}
\label{fig:shower45}
\end{figure}

\begin{figure}
\centering
\includegraphics[width=0.95\textwidth]{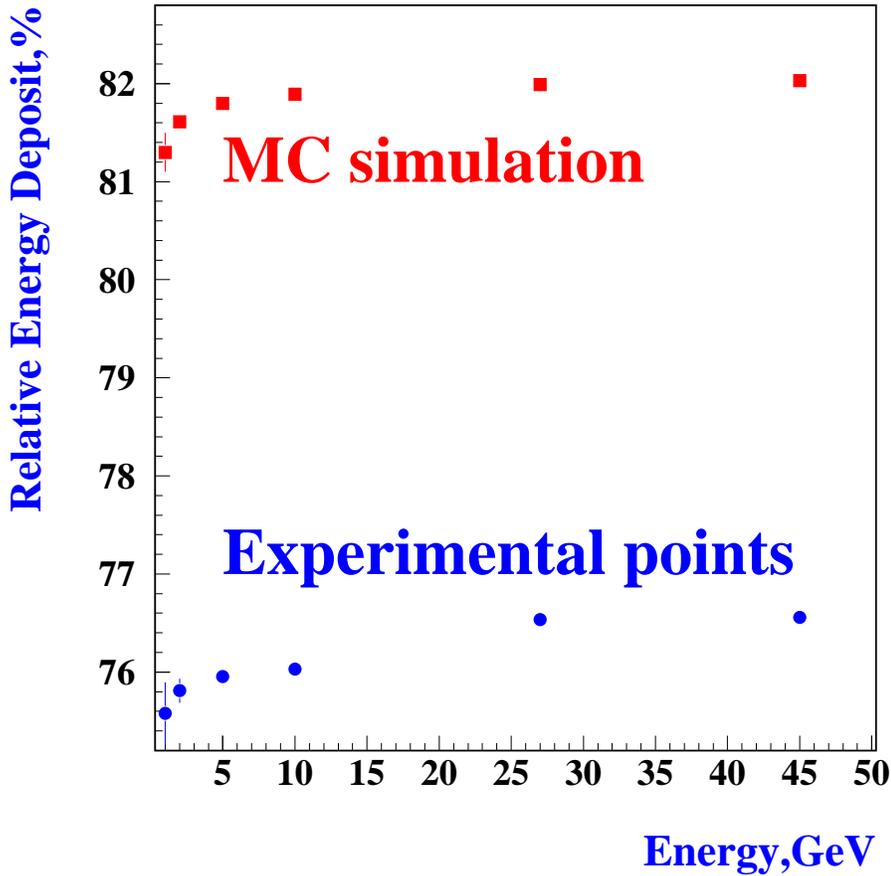}
\caption{The dependence of the energy deposit in the central crystal of the
$5\times 5$ matrix on energy. Electrons hit a square of $4\times 4$ mm$^2$ 
in the center of the matrix.
}
\label{fig:center_eloss}
\end{figure}

A typical electromagnetic shower profile, for 45 GeV electrons, is presented
in Fig.~\ref{fig:shower45}. It defines an average energy deposit, normalized
on the total energy deposit in the 5$\times$5 crystal array, in the
crystal with a distance $X$ between its center and an electron
$X$-coordinate. The electron $Y$-coordinate here is within 1~mm
slice of the crystal center.
We found that the lateral shower profile changes very slightly
in the energy range 1--45 GeV. When an electron hits the square 
of 4$\times$4 mm$^2$ in the center of middle crystal of the calorimeter
prototype, about 76\% of the full energy  is deposited in this
crystal. The dependence of the energy deposit on energy is 
presented in Fig.~\ref{fig:center_eloss}.
In Table \ref{tab:edep} the relative energy deposit (in \%) in the
six crystals when an electron hits the square of 4$\times$4 mm$^2$ in 
the center of the matrix is presented. These six counters 
reproduce the triangle matrix obtained as described above when 
each cell of the arrays corresponds to a crystal.
The upper lines of the numbers stand for an
average over 1 and 2 GeV, while the bottom lines show values for 45 GeV.
We see that while the energy changes by about a factor of 30, the shower
profiles changes by only 1\% in the center and at most 10\% at the edges.
  
\begin{table}
\begin{center}
\caption{Energy deposit in \% when an electron hits 
the center of the crystal matrix. The upper lines are for an average of 1 and 2 GeV and the bottom lines for 45 GeV.}
\begin{tabular}{|r|c|c|c|}
\hline
       &center & center+1 & center+2\\
\hline
center, 1-2 GeV & 75.7 & & \\
       45 GeV& 76.6 & & \\ 
\hline
center+1, 1-2 GeV & 3.92  &  1.180  & \\
         45 GeV &  3.83    &  1.137  & \\
\hline
center+2, 1-2 GeV & 0.402 &  0.250 &   0.076 \\  
         45GeV &  0.375   &  0.226   & 0.069  \\
\hline
\end{tabular}
\label{tab:edep}
\end{center}
\end{table}

     The test beam results on the electromagnetic shower lateral
profile have been compared with the GEANT 3.21 simulation results
with tracing cuts of 500~KeV for electrons/positrons and 60~KeV
for photons. The GEANT 4 simulation with the same tracing cuts and
with the much lower 1~KeV production and tracing cuts gives
very similar results. The GEANT simulation indicates that the
simulated shower transverse profile does not change with
the incoming energy. About 82\% of the energy collected over
$5\times 5$ matrix is predicted by GEANT to be in the central crystal, but
our measurements show only about 76-77\%. GEANT does better when considering 
the fraction of the energy contained in the $3\times3$ sub-array; GEANT 
predicts 97\%, while the measurement is 96\%. It is also interesting that 
the shower profile simulated with
GEANT is a bit narrower in comparison to the experimental result. One of the
possible explanations may be Cherenkov's effect light contribution.

\section{Position Resolution and Comparison With GEANT Simulations}

\begin{figure}
\centering
\includegraphics[width=0.95\textwidth]{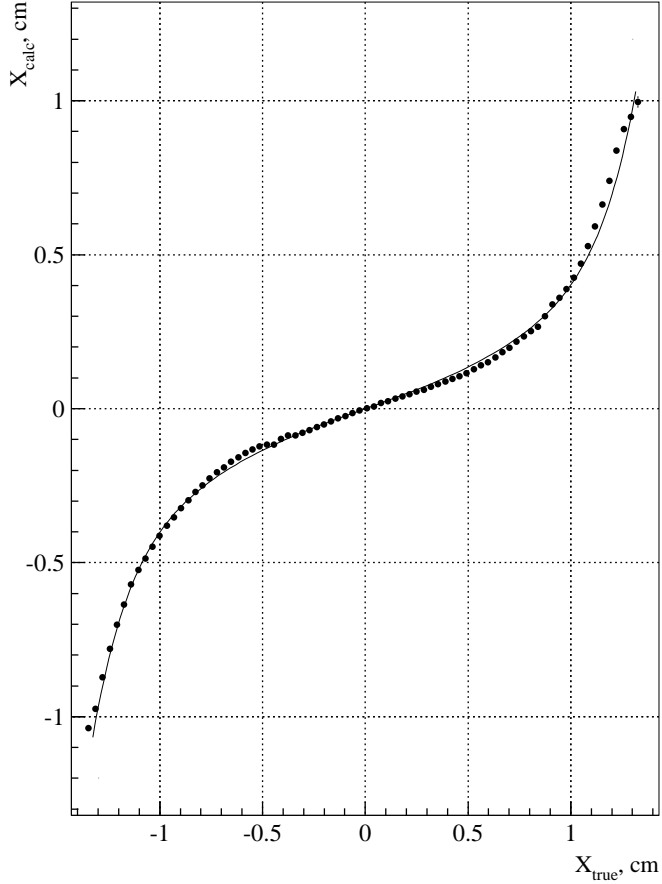}
\caption{ Measured position versus true position defined by the 
drift chambers  (S-curve).}
\label{fig:s_wave}
\end{figure}

\begin{figure}
\centering
\includegraphics[width=0.95\textwidth]{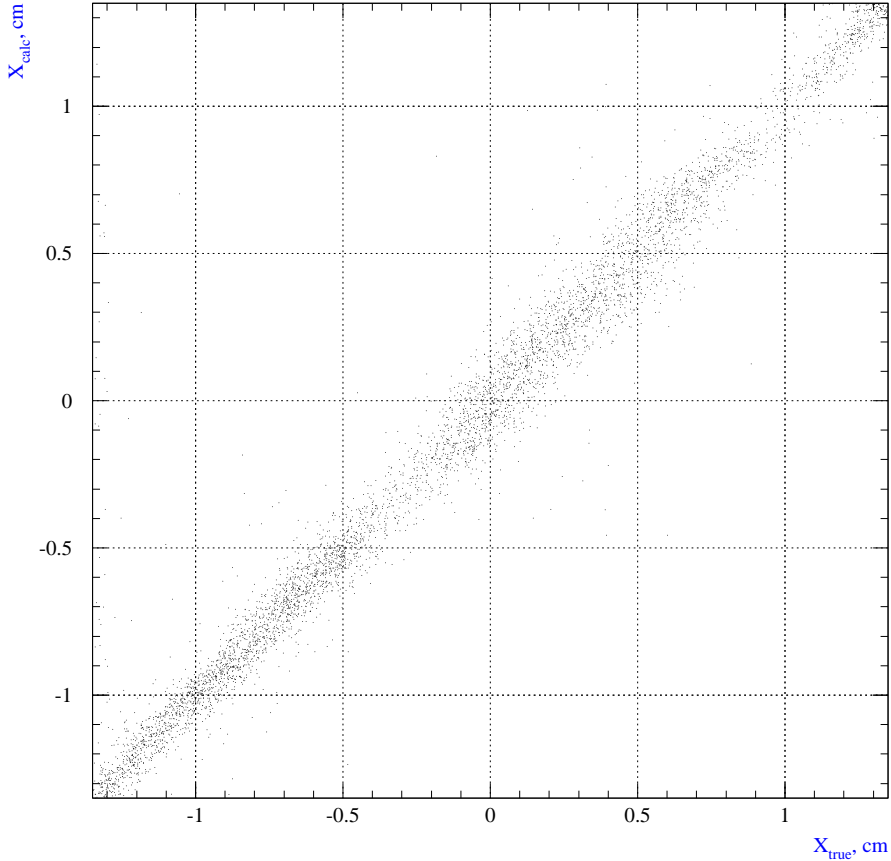}
\caption{Dependence of the calculated coordinate (along the y-axis) in 
the PbWO$_4$ matrix on the true coordinate (along the x-axis)
defined with the use of the drift chambers.
}
\label{fig:calc_true}
\end{figure}

      The position resolution of the calorimeter prototype  was also obtained 
using the test beam data. We used the center-of-gravity technique to
define the measured coordinate in the crystal matrix and the drift
chambers to define the true coordinate of the particle hitting
the crystals. This method requires the knowledge of 
the energy and the angle of incidence of the particle in order 
to choose the right correction curve (so called, the S-curve) which 
will be used to correct the measured position. The measured position 
$x_{meas}$ is defined as follows:

\begin{equation}
      x_{meas} = \sum_{i=1}^{n}{E_i\cdot x_i}/\sum_{i=1}^{n}{E_i}~,
\label{eq:x_meas}
\end{equation}
where $E_i$ is the energy deposited in the i-th
crystal and $x_i$ is the position of its center, relative to (0,0) for the central crystal, and $n$ refers to the number of crystals in the sum. 
While for the energy resolution it is better to take $n=25$, for the
spatial resolution {\it n} prefers to be taken as 9 due to the influence
of large energy fluctuations in the tails of an electromagnetic shower
on the spatial resolution. The measured position versus 
the true position
for 27 Gev electrons for normal incidence is presented in Fig.~\ref{fig:s_wave}
(S-curve). The weighted sum (\ref{eq:x_meas}) biases 
the measurement towards 
the center of the crystal as can been seen in this figure. For normal 
incidence the correction S-curve is almost independent 
of energy, most likely due to the fact that the Moliere radius has a negligible
energy dependence.
The S-curve fit function was used to obtain a calculated coordinate. 
The dependence of the calculated coordinate 
in the PbWO$_4$ matrix on the true coordinate defined by the drift 
chambers for 27 GeV electrons is presented in 
Fig.~\ref{fig:calc_true}. The width of the line in 
Fig.~\ref{fig:calc_true} gives the calorimeter prototype 
position resolution. 
     
\begin{figure}
\centering
\includegraphics[height=0.4\textheight]{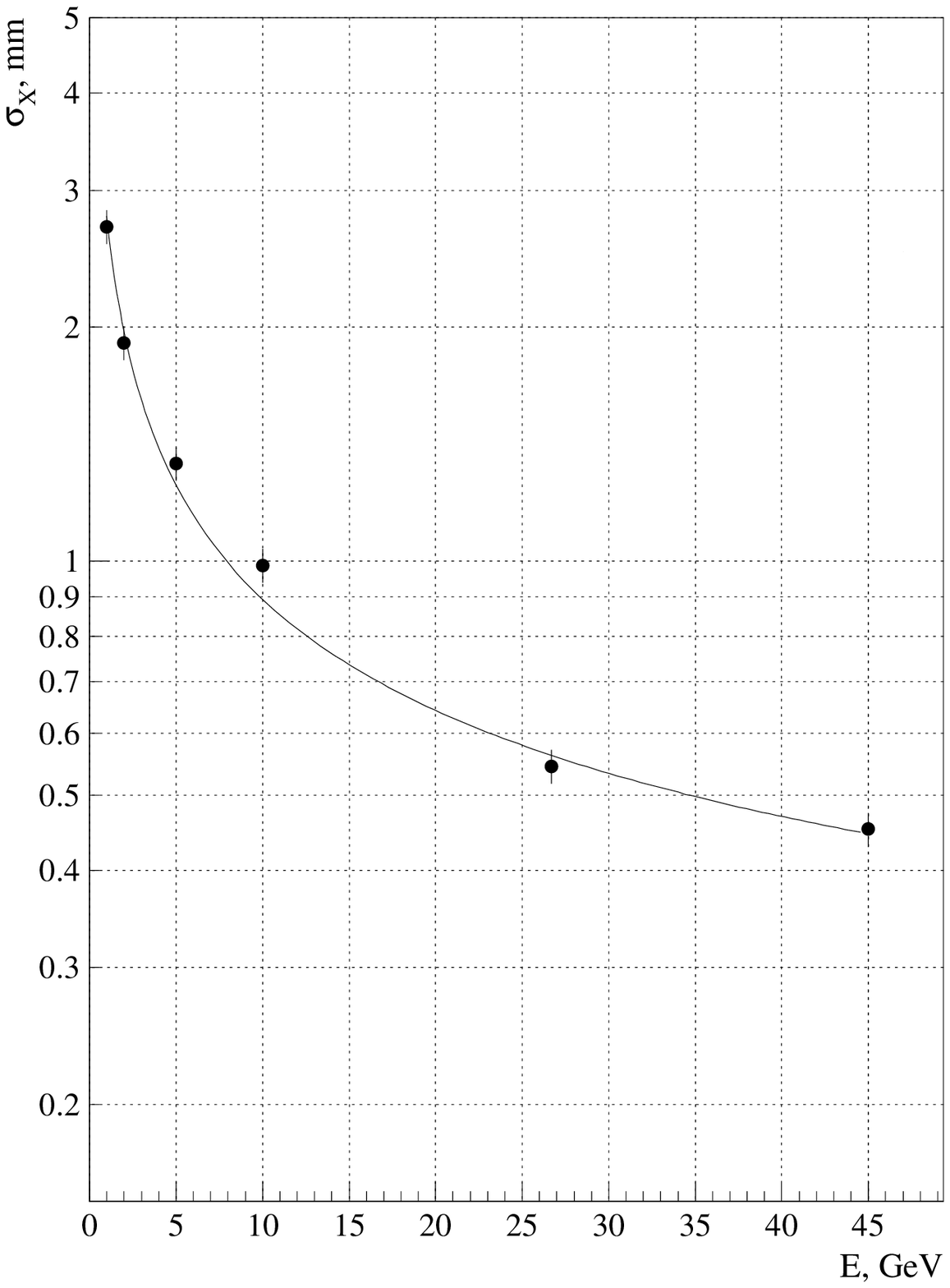}
\caption{Measured position resolution dependence on energy 
of the $3\times 3$ crystal matrix. Curve represents a fit 
of the experimental data. 
}
\label{fig:sigma_x_E}
\end{figure}

     The resolution, averaged over electrons spread across the entire
central crystal was calculated for several beam energies. The results
are  shown in Fig.~\ref{fig:sigma_x_E}. We obtained the following 
dependence of resolution on energy 
\begin{equation}
\sigma_x = (0.16 \pm 0.06) \oplus \frac{2.80 \pm 0.08}{\sqrt{E}}~, 
\end{equation}
({\it E} in units of GeV and $\sigma_x$ in mm).
This agrees well with the resolution expected from Monte Carlo simulation,
which is
\begin{equation}
\sigma_x = (0.17 \pm 0.01) \oplus \frac{2.77 \pm 0.01}{\sqrt{E}}~. 
\end{equation}

\begin{figure}
\centering
\includegraphics[height=0.4\textheight]{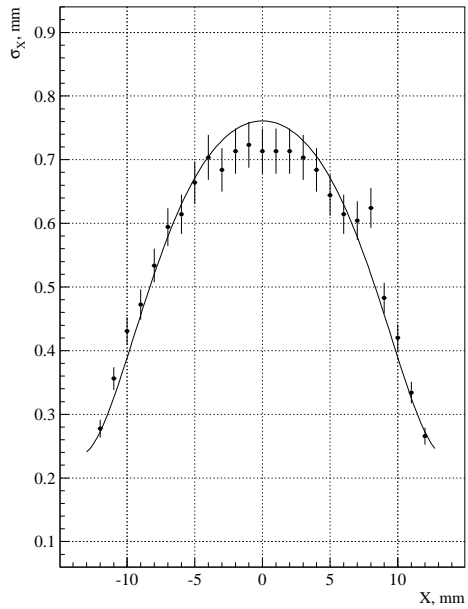}
\caption{Measured dependence of the spatial resolution on the position of 
a beam electron inside the crystal. Zero on the X-axis is the center 
of the crystal. The curve is the result 
of Monte Carlo simulation.
}
\label{fig:sigma_x_X}
\end{figure}

The spatial resolution strongly depends on the position of 
an electron hitting the matrix relative to
the center of a crystal, as shown in Fig.~\ref{fig:sigma_x_X}. 
We see that the spatial resolution is better by almost a factor of three
the edge of the crystal relative to the center. 
As shown in Figure~\ref{fig:sigma_x_X}, experimental resolution in the 
center of the crystall is better than MC simulation, since 
GEANT simulated shower is a bit narrower than measured.

\begin{figure}
\centering
\includegraphics[width=0.7\textwidth]{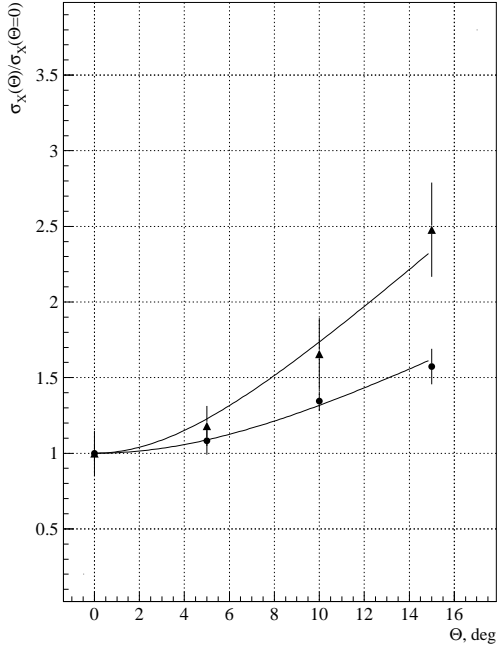}
\caption{The dependence of the position resolution on the angle 
normalized on the position resolution at zero degree. The filled 
circles resulted from 10 GeV electrons, and the triangular points 
from 27 GeV. The curves are the results of a fit described in the text.  }
\label{fig:coord_angle}
\end{figure}

    In order to study the dependence of position resolution on the angle 
at which
an electron hits the calorimeter prototype, the crystal matrix was 
turned by 5, 10 and 15 degrees relative to normal.
Data were taken at 10 and 27 GeV for each angle. 
At each angle the S-curve was determined and fitted. 
The dependence of position resolution
on angle relative to the one at zero degree incidence for 10 GeV and 27 GeV electrons 
is given in Fig.~\ref{fig:coord_angle}. We see that position resolution at 
15 degrees is worse than at zero degree by a factor of 1.5 at  
10 GeV and 2.5 at 27 GeV. The curves in the figure
are results of fits by the function:
\begin{equation}
\sigma_{\theta}/\sigma_0 = 1 \oplus A \cdot \sin{\theta}, 
\end{equation}
where $\sigma_{\theta}$ is a position resolution at angle $\theta$, 
$\sigma_0$ is a position resolution at zero degree, 
$A$ -- fit parameter.
When the matrix is
turned at some angle, the additional contribution in position resolution
comes from longitudinal fluctuations of the electromagnetic shower maximum.
The geometrical factor of this effect is proportional to $\sin{\theta}$.

\section{Conclusion}

     The measurements of energy and position resolutions of 
Our electromagnetic calorimeter prototype made of lead
tungstate crystals for the BTeV experiment at Fermilab have been
carried out at the IHEP test beam facility at the
Protvino 70 GeV accelerator. The crystals were produced in
Bogoroditsk (Russia) and Shanghai (China) and
were assembled in 5x5 arrays.  

Studies were made in the electron beam energy range from 1 to 45
GeV. The energy tagged beam has allowed us to measure the stochastic
term in energy resolution as (1.8$\pm$0.1)\%. 
We have not seen significant difference in energy resolution of the
Bogoroditsk and Shanghai crystals.
The non-uniformity of light yield along the crystal has been
measured with the use of the muon beam when the crystal matrix was
rotated by 90 degrees with respect to the beam direction. Taking
into account this effect as well as photostatistics has resulted
in good agreement between the measured energy resolutions  and
the GEANT Monte Carlo simulations.

      The stochastic term in the dependence of position resolution
on energy in our measurements is about 2.8~mm which is in agreement
with Monte Carlo simulations. For 27 GeV electrons position resolution is
750 $\mu$m in the center of the crystal, and is 250 $\mu$m at a
boundary between two crystals.

      The dependence of energy resolution
on the angle at which particle hits the crystals relative to the
normal has been measured. Energy resolution does not deteriorate
until the angle is about 5$^{\circ}$. But there is significant 
dependence of position resolution on the angle. The position resolution at 
15$^{\circ}$ is worse than at 0$^{\circ}$ by a factor of 1.5 at 10 GeV and 2.5 
times for 27 GeV. Projective geometry rather than planar geometry has been 
chosen for the BTeV calorimeter. 

      The electromagnetic shower lateral profile changes very slightly
in the energy range 1--45 GeV. When an electron hits the center of
the calorimeter prototype made of a PbWO$_4$ with the size
$27\times 27\times 220$ mm$^3$, about 76\% of the full energy is 
deposited in the central crystal. Shower shape tables were measured
for six energies
within 1--45 GeV region. It will be important for developing an
algorithm of a reconstruction of two gamma-quanta from two
overlapped showers at different energies.
The shower profile simulated with
GEANT is a bit narrower in comparison to the experimental result.

      The temperature dependence of a crystal light output was measured to be
-2.3\%/$^{\circ}$C (at 18~$^{\circ}$C) for two electron beam energies, 
10 and 27 GeV.

\section{Acknowledgments}
    We would like to thank the IHEP management for providing us
a beam line and accelerator time for our test beam studies.
Special thanks to Fermilab for providing
equipment for data acquisition.The authors would like to 
thank O.A.~Grachov and I.V.~Kotov for useful discussions. 
This work was partially supported by the U.S.
National Science Foundation and the Department of Energy.

\end{document}